\documentclass[aps,prm,reprint, superscriptaddress]{revtex4-2}
\usepackage{blindtext}
\usepackage{graphicx}
\usepackage{xcolor}
\usepackage{dcolumn}
\usepackage{bm}
\usepackage[utf8]{inputenc}

\begin{document}
	
	\preprint{APS/123-QED}
	
	\title{Spontaneous Stabilization of a Hidden-Phase-Like Metallic State in 1T-TaS$_2$}

	\author{Anil Rajapitamahuni}
	\email{s-arajapi1@unl.edu }
	\affiliation{Department of Physics, SRM University - AP, Amaravati, Andhra Pradesh, 522502,India}
	\affiliation{Department of Physics and Astronomy, University of Nebraska-Lincoln, Nebraska 68588, USA}
	
	\author{Suji Park}
	\affiliation{Center for Functional Nanomaterials,  Brookhaven National Lab, Upton, New York 11973, USA}

	\author{Houk Jang}
	\affiliation{Center for Functional Nanomaterials,  Brookhaven National Lab, Upton, New York 11973, USA}
	
	\author{Asish K. Kundu}
	\affiliation{National Synchrotron Light Source II, Brookhaven National Lab, Upton, New York 11973, USA}
	
	\author{Elio Vescovo}
	\affiliation{National Synchrotron Light Source II, Brookhaven National Lab, Upton, New York 11973, USA}
	
	\author{Turgut Yilmaz}
	\email{trgt2112@gmail.com}
	\affiliation{Department of Physics, Xiamen University Malaysia, Sepang 43900, Malaysia}
	\affiliation{Department of Physics, University of Connecticut, Storrs, CT 06269, USA}

\date{\today}

\begin{abstract}
Electronic phases that lie outside the equilibrium ground state offer a route to explore competing configurations in correlated materials. In 1T-TaS$_2$, ultrafast excitation accesses a metallic hidden phase that is distinct from the commensurate insulating ground state. Here we use angle-resolved photoemission spectroscopy to show that a hidden-phase-like electronic configuration can be spontaneously stabilized in exfoliated intermediate-thickness 1T-TaS$_2$ flakes, where it persists up to room temperature before evolving through a different sequence of electronic transitions. This hidden-phase-like metallic state hosts a metallic band with finite Fermi‑level spectral weight while retaining the characteristic hybridization gaps associated with the star‑of‑David band folding. These results establish a platform for controlling competing electronic states in layered materials, with implications for both quantum science and phase change technologies.

\end{abstract}

	\maketitle
	
	\section{Introduction}
	
     Stabilizing fragile or metastable electronic phases in bulk is central to the discovery of new materials and to advancing our understanding of correlated quantum matter~\cite{bao2022light,basov2017towards}. Hidden phases, electronic states inaccessible under equilibrium yet revealed by external perturbations such as ultrafast excitation, strain, or confinement, represent a particularly intriguing frontier~\cite{domrose2023light,gao2022snapshots}. Demonstrating that such phases can become accessible without external optical or electrical stimulation expands the accessible phase space and provides a route toward engineering new electronic functionalities.

	1T-TaS$_2$ provides a paradigmatic platform for exploring hidden phases in correlated quantum materials. In bulk, it undergoes a cascade of charge density wave (CDW) transitions: from a commensurate (C-CDW) state below $\sim$180~K, to a nearly commensurate CDW (NC-CDW) between $\sim$240$-$350~K, then to an incommensurate CDW (I-CDW) above $\sim$350~K, before transforming into a metallic phase near $\sim$550~K~\cite{sipos2008mott}. The C-CDW is characterized by a sharp resistivity jump and an emergent energy gap. The origin of this insulating ground state remains under active debate, with interpretations ranging from a Mott insulator stabilized by CDW-induced band narrowing~\cite{sipos2008mott} to a band insulator driven by interlayer dimerization and stacking order~\cite{wang2020band}. An alternative scenario has been also suggested that stacking-dependent interlayer hybridization is a key ingredient in the formation of the low-temperature insulating ground state~\cite{ritschel2015orbital}. Beyond this equilibrium sequence, ultrafast excitation can transiently induce a metastable metallic phase known as hidden phase~\cite{stojchevska2014ultrafast,vaskivskyi2015controlling,vaskivskyi2016fast,huber2025revealing}.

	Spectroscopically, the hidden phase is characterized by the emergence of a shallow, dispersive electron-like band centered at the Brillouin-zone center, accompanied by a collapse of the lower Hubbard band~\cite{huber2025revealing,maklar2023coherent}. This phase retains CDW distortions, which give rise to weak residual band‑folding features, although the long‑range CDW coherence is reduced compared to the commensurate state. The metallic band (MB) carries finite spectral weight at the Fermi level and is absent in the commensurate CDW/Mott state, where the electronic structure is dominated by a flat lower Hubbard band and a full gap at $E_F$. The hidden phase therefore represents a distinct electronic configuration.

	Stabilizing a hidden-phase-like electronic configuration without external stimulation could offer new insight into how competing electronic orders reshape the underlying electronic landscape. To date, no spectroscopic study has demonstrated a hidden-phase-like electronic structure in exfoliated 1T-TaS$_2$ flakes under equilibrium conditions. Notably, scanning tunneling studies have shown that, within small domains, a metallic mosaic phase can spontaneously emerge in equilibrium due to surface defects or stacking variations~\cite{ma2016metallic,salzmann2023observation}, suggesting that hidden phase may be stabilized under specific structural conditions.

	\begin{figure*}[t]
		\centering
		\includegraphics[width=0.9\textwidth]{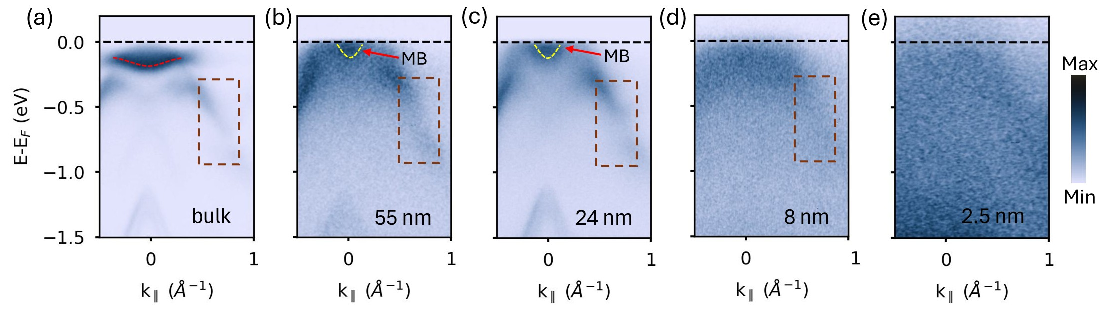}
		\caption{\textbf{Dimensional control of the electronic structure in 1T-TaS$_2$.} 
			(a-e) ARPES spectra along $\Gamma$--M are shown for bulk and exfoliated flakes with thicknesses of 55, 24, 8, and 2.5~nm, respectively. Bulk crystals exhibit the C-CDW state with a lower Hubbard band at $\approx$ $\sim$200~meV marked with red parabola with no spectral weight at $E_F$. By contrast, 24 and 55~nm flakes display a shallow $\Gamma$-centered MB crossing $E_F$ marked with yellow dashed parabolas, consistent with stabilization of the HPLMS under equilibrium conditions. Thinner flakes show progressively degraded spectra: the 8~nm sample retains a broadened MB, while the 2.5~nm flake evolves into a fully gapped state with incoherent features marked with dashed black line. Dashed brown rectangles mark representative hybridization gaps from CDW band folding. All samples exhibit similar hybridization gaps except the 2.5~nm flake. All ARPES data is recorded at 50 K sample temperature with 92 eV photon energies.
		}
		\label{fig:thickness}
	\end{figure*}

Exfoliated flakes of 1T-TaS$_2$ provide a useful platform to explore this question. Transport studies show that while flakes remain insulating at low temperatures, the sharp bulk-like Mott transition near 240~K is absent. Instead, a broad crossover near 350~K is observed, suggesting that reduced thickness fundamentally alters the competition between insulating and metallic states~\cite{yoshida2015memristive,yoshida2014controlling}. However, direct spectroscopic evidence linking these anomalous transport features to the underlying electronic structure has remained lacking. Here, we use high-resolution angle-resolved photoemission spectroscopy (ARPES) to investigate exfoliated 1T-TaS$_2$ over a wide temperature range. We find that thin flakes stabilize a hidden-phase–like metallic state (HPLMS) with finite spectral weight at the Fermi level from 50~K to 300~K. Notably, this stabilization extends to flakes tens of nanometres thick, highlighting the sensitivity of the correlated state to modest structural variations. Crucially, our temperature‑dependent ARPES reveals that the electronic structure circumvents the sharp 240~K Mott transition and instead undergoes a two‑step crossover: CDW coherence weakens above 270~K and quasiparticle coherence collapses near 370-380~K. This spectroscopic trajectory with the absence of a sudden gap opening directly explains the transport anomalies of the flakes.

\begin{figure*}[t]
	\centering
	\includegraphics[width=0.75\textwidth]{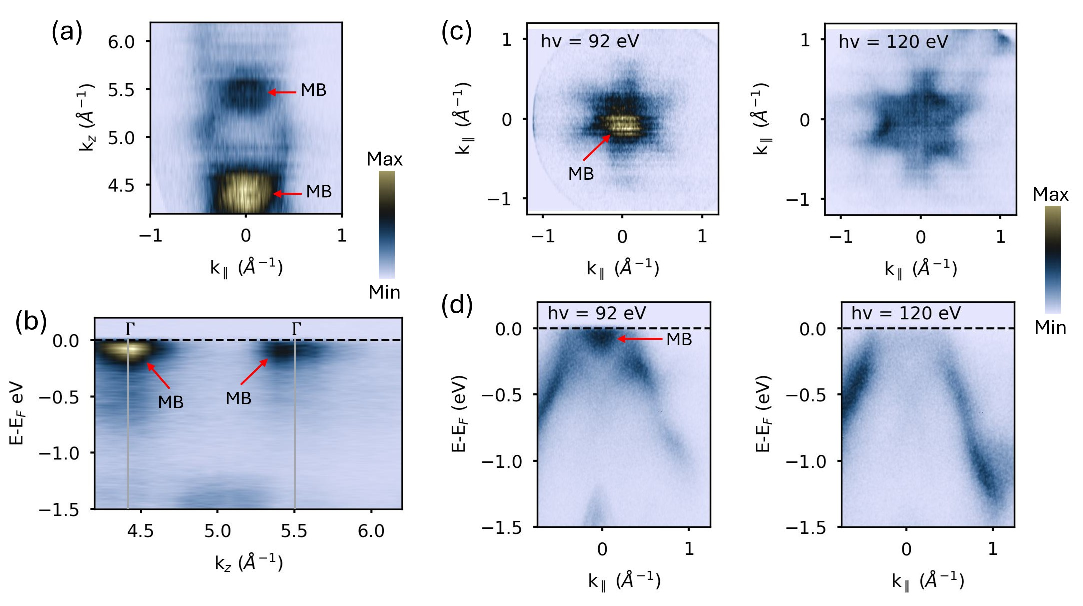}
	\caption{\textbf{$k_z$ periodicity and photon-energy dependence of the MB in exfoliated 24 nm thick 1T-TaS$_2$.}  
		\textbf{a}, Fermi surface map in the ($k_\parallel$, $k_z$) plane showing strong intensity of the MB only at selected $k_z$ values.  
		\textbf{b}, Corresponding $E$--$k_z$ dispersion highlighting the $\Gamma$-centered MB.  
		\textbf{c}, Fermi surface maps at $h\nu = 92$~eV ($k_z \approx \Gamma$) and 120~eV ($k_z \approx A$), showing a bright pocket at $\Gamma$ that vanishes at $A$.  
		\textbf{d},  ARPES spectra measured along the $\Gamma$–$M$ and $A$–$L$ directions at the same photon energies, reinforcing the photon-energy dependent modulation of MB. All ARPES data is recorded at 50 K sample temperature.
	}
	\label{fig:kz}
\end{figure*}

	\section{Thickness-dependent stabilization of the HPLMS}
	
    Thin flakes of 1T-TaS$_2$ were mechanically exfoliated inside an inert-atmosphere glove box. Details of the exfoliation procedure, flake characterization, and transfer to the ARPES beamline are provided in the Supplemental Material~\cite{SM}, together with representative optical images (Fig.~S1) and core-level spectra (Fig.~S2). We first examine the electronic structure of exfoliated 1T-TaS$_2$ flakes using systematic ARPES measurements across thicknesses ranging from 2.5 to 55 nm, compared with the one of bulk crystals. In the bulk, consistent with previous studies~\cite{yang2022visualization}, the spectra exhibit a relatively flat band located ~200 meV below the Fermi level, leaving an insulating gap. This feature is commonly attributed to the lower Hubbard band~\cite{dardel1992temperature,perfetti2005unexpected,butler2020mottness}. Multiple folded replicas arising from the star-of-David CDW reconstruction induce hybridization gaps, which are visible at higher binding energies and marked by the dashed brown rectangle in Fig. 1a.

	Unlike the bulk sample, 55 and 24~nm flakes host a well-defined $\Gamma$-centered MB at the $E_F$. Its dispersion and finite Fermi-level spectral weight closely mirror the hidden phase previously induced through ultrafast excitation or current~\cite{maklar2023coherent,huber2025revealing}, but here it emerges spontaneously in exfoliated flake samples without the need for optical or electrical excitation. Importantly, this MB coexists with residual CDW fingerprints: the spectra retain characteristic suppression of intensity at selected energy–momentum points, arising from hybridization with folded bands. These observations suggest, from a spectroscopic perspective, that the HPLMS can emerge spontaneously in the exfoliated flakes.
	
	At reduced thicknesses, however, spectral coherence diminishes. In 8~nm flakes, the MB remains visible but is substantially broadened, with incoherent weight spreading to higher binding energies and diminished quasiparticle intensity at $E_F$. In 2.5~nm flakes, spectral weight near the Fermi level is even further suppressed and no CDW folding can be resolved. This behavior is consistent with transport measurements, which reveal either a strongly weakened or completely absent resistivity anomaly in thinner flakes~\cite{yoshida2015memristive,yoshida2014controlling}, emphasizing the fragility of coherent quasiparticles in the exfoliated samples.

	To further clarify the low-energy electronic structure, representative EDCs for all investigated thicknesses are provided in Fig.~S3 \cite{SM}. The EDCs confirm the emergence of enhanced spectral weight near the Fermi level at the $\Gamma$ point in the 55~nm and 24~nm flakes, consistent with the metallic band discussed above. In contrast, the hole-like valence band at $k_{||}=0.36$~\AA$^{-1}$ remains separated from the Fermi level for all investigated thicknesses, indicating that it does not cross the Fermi level.
	
	\section{$k_z$-resolved electronic structure of the HPLMS}  
	
    To probe the three-dimensional nature of the HPLMS, we examined the out-of-plane electronic structure using photon-energy-dependent ARPES. This approach reveals how electronic states evolve with $k_z$, providing a decisive test of whether the stabilized metallic state is a bulk-derived feature or a purely two-dimensional surface effect.

    The MB exhibits a pronounced and selective $k_z$ dependence. As shown in the Fermi surface map of the $(k_{\parallel}, k_z)$ plane (Fig.~2a), the MB carries strong spectral weight only at specific $k_z$ values near the Brillouin zone center ($k_z \approx \Gamma$), with its intensity rapidly diminishing elsewhere. This selective presence is a characteristic spectroscopic fingerprint of the HPLMS and demonstrates that the metallic band possesses a pronounced three-dimensional character. The observed photon-energy dependence indicates substantial interlayer coherence, inconsistent with a purely surface-localized metallic state.
    
    The $k_z$-confinement of the MB is further illustrated in Fig.~2c,d. At $h\nu = 92$ eV (corresponding to $k_z \approx \Gamma$), a well-defined, $\Gamma$-centered electron pocket is visible in the Fermi surface and as a shallow band crossing $E_F$ in the dispersion. In contrast, at $h\nu = 120$ eV (corresponding to $k_z \approx A$, the zone boundary along $c^*$), this pocket vanishes. The corresponding energy-momentum dispersion (Fig.~2d, right panel) shows that the coherent MB is absent at this $k_z$. This stark contrast confirms that the HPLMS is not uniformly metallic in all three dimensions but is stabilized within a specific region of the bulk Brillouin zone, indicative of a modified three-dimensional coherent electronic state.
    
    The periodic modulation of the MB intensity further enables a direct measurement of the $c$-axis lattice parameter. Within the free-electron final-state approximation, the modulation period yields $c = 2\pi/\Delta k_z \approx 5.86~\text{\AA}$, in excellent agreement with the bulk value~\cite{hu2018toward,cao2020complete}. This confirms that the exfoliated flakes preserve the long-range stacking periodicity of the crystal lattice. Together, these results demonstrate that intermediate-thickness flakes maintain the bulk crystallographic framework while hosting a spontaneously stabilized HPLMS, a state with a distinct, $k_z$-selective metallic band that underscores the sensitivity of the correlated ground state to structural confinement.

    \begin{figure*}[t]
    	\centering
    	\includegraphics[width=0.85\textwidth]{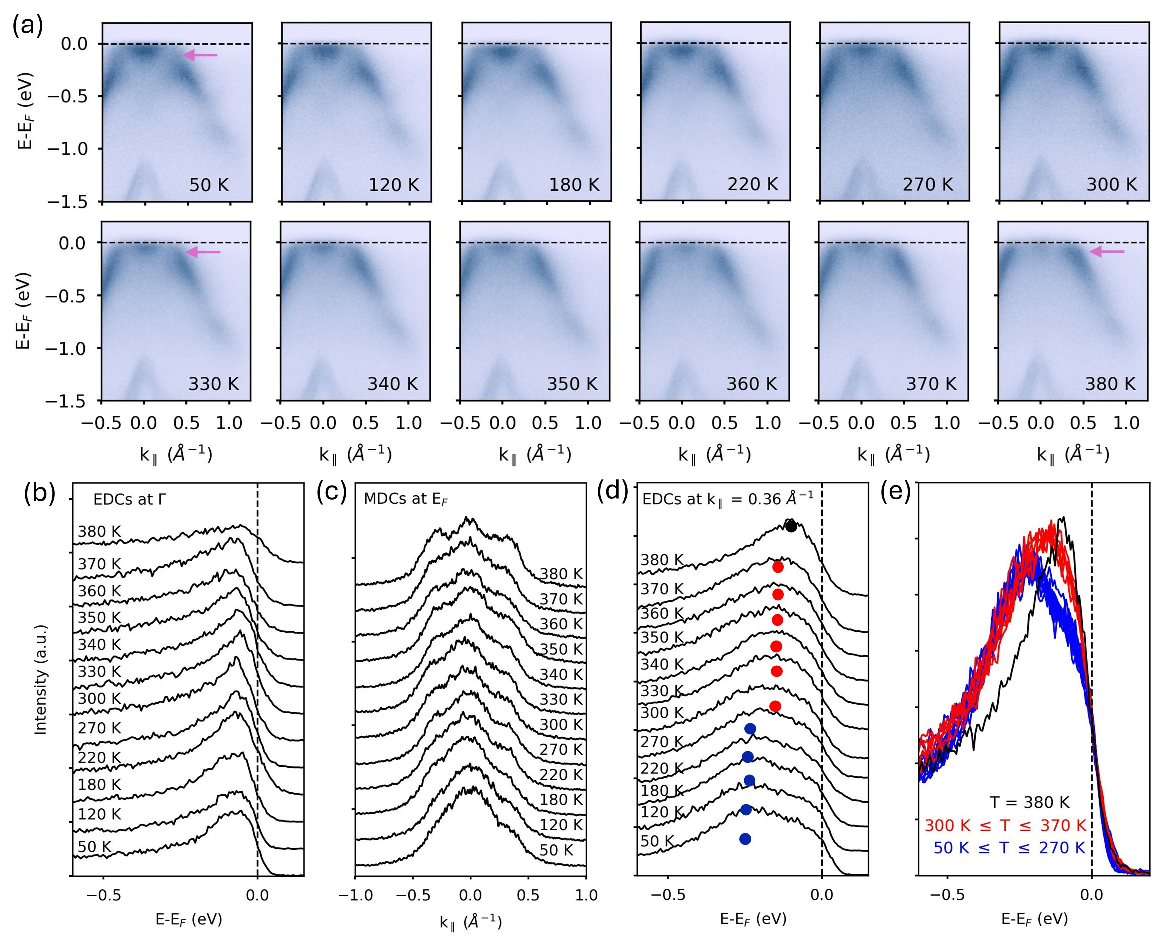}
    	\caption{\textbf{Temperature-dependent ARPES of a 24~nm 1T-TaS$_2$ flake.}  
    		\textbf{a}, ARPES spectra along $\Gamma$–M recorded between 50 and 380~K. At low $T$, the MB coexists with a CDW-related suppression near 0.2–0.3~eV below $E_F$, persisting up to 300~K. Above 300~K, the suppression weakens, and at 370–380~K the CDW feature collapses, leaving only a broadened MB. The magenta-colored arrows indicate the regions of suppressed spectral weight associated with CDW induced hybridization.
    		\textbf{b}, Energy distribution curves (EDCs) at $\Gamma$ showing a sharp loss of MB spectral weight across 370~K.  
    		\textbf{c}, Momentum distribution curves (MDCs) at $E_F$, integrated within 30~meV below $E_F$, evolving from a broad MB-dominated profile at low $T$ to distinct side-band peaks at high $T$.  
    		\textbf{d}, EDCs at $k_\parallel = 0.36~\text{\AA}^{-1}$, where the CDW-related peak at $\sim$0.23~eV (blue markers) is suppressed with increasing $T$, while new spectral weight emerges near $E_F$ (red markers).    
    		\textbf{e}, Scaled EDCs extracted at $k_\parallel = 0.36~\text{\AA}^{-1}$ and integrated within $0.05~\text{\AA}^{-1}$, tracking the temperature evolution of the CDW-related spectral feature. The colored grouping highlights the gradual spectral evolution above $\sim$270--300~K and the pronounced change near 370--380~K.
    	}
    	\label{fig:temperature}
    \end{figure*}

	\section{Temperature-dependent electronic structure of exfoliated flakes}  
	
	Transport studies on exfoliated 1T-TaS$_2$ flakes show that the bulk 180--240~K metal--insulator transition is suppressed, with only a crossover near 350--370~K~\cite{yoshida2014controlling,yoshida2015memristive}. This altered sequence of transitions raises two key questions: what replaces the commensurate CDW/Mott phase in intermediate-thickness flakes, and how stable is the HPLMS with temperature?  
	
	Fig.~3 shows ARPES spectra of a 24~nm flake measured from 50 to 380~K. At low $T$, the MB coexists with CDW-related suppression near $E_F$ that persists up to 270 K. Notably, no bulk-like insulating gap appears in the 180–240~K range, in contrast to bulk crystals where a full gap opens at $\Gamma$~\cite{wang2020band,dardel1992temperature}. Above 300~K, CDW-related suppression weakens while the overall band structure remains intact until 370~K, where the CDW feature collapses and the MB is reduced to a broadened remnant at 380~K. The temperature evolution of the electronic structure in flakes is also consistent with the behavior of the metastable state induced in bulk samples through the current~\cite{huber2025revealing}, further confirming consistence of our data with the previous reports.

	This two-step redistribution of spectral weight (evident in the scaled EDCs of Fig. 3e) explains the transport behavior of flakes: gradual suppression of the CDW peak between 300 and 370~K corresponds to the smooth resistivity slope, while the abrupt collapse near 370–380~K accounts for the resistivity anomaly. It should be noted that the transport response of exfoliated 1T-TaS$_2$ flakes is strongly dependent on the cooling and heating rate. At sufficiently slow cooling ($\approx 0.2~\mathrm{K\,min^{-1}}$), a Mott-like transition can re-emerge in flakes tens of nanometres thick, whereas this feature is absent under faster cooling process~\cite{yoshida2014controlling}. In our ARPES measurements, the flakes and the bulk sample are transferred onto a cold sample stage, resulting in rapid cooling and thus placing the samples in the fast-cooled regime.

\section*{Discussion and Conclusion}

In summary, bulk crystals cleaved \textit{in situ} exhibit the well-established C-CDW phase with a gapped low-temperature electronic structure, whereas exfoliated flakes develop a $\Gamma$-centered metallic band (MB) without the bulk insulating gap. In the few-layer limit, the spectra evolve toward an incoherent gapped state. This evolution closely mirrors thickness-dependent transport measurements, which likewise report suppression of the insulating transition in intermediate-thickness flakes and recovery of gapped behavior in the ultrathin limit. The metallic state remains stable from cryogenic temperatures to room temperature and exhibits a two-step thermal evolution: partial weakening of CDW-related spectral features above $\sim$270~K, followed by a collapse of quasiparticle coherence near 370--380~K. These characteristic temperatures closely correspond to the anomalous transport behavior reported for intermediate-thickness flakes, establishing a direct spectroscopic link to their unusual conductivity.

A notable feature is the non-monotonic thickness dependence. While intermediate-thickness flakes exhibit the HPLMS, the metallic state is absent in both bulk crystals and the few-layer limit. This behavior is consistent with previous transport studies showing that reducing the thickness suppresses the NCCDW--CCDW transition, stabilizing a supercooled NCCDW state only within an intermediate thickness range \cite{yoshida2014controlling,yoshida2015memristive}. In this context, the HPLMS may represent the spectroscopic manifestation of the anomalous electronic state that emerges when the bulk NCCDW--CCDW transition is suppressed.

The microscopic mechanism responsible for stabilizing the HPLMS remains to be established. Previous studies have shown that hidden and metastable states in 1T-TaS$_2$ are characterized by suppression of the insulating ground state and the emergence of metallic electronic states accompanied by modifications of the charge order \cite{stojchevska2014ultrafast,vaskivskyi2016fast}. STM measurements have further revealed equilibrium metallic mosaic states \cite{salzmann2023observation,ma2016metallic}, while ARPES studies of current-induced hidden states reported a $\Gamma$-centered metallic band with finite spectral weight at the Fermi level \cite{huber2025revealing,nitzav2026emergence}. The close similarity between these spectroscopic signatures and those observed here suggests that the HPLMS arises from a related electronic reconstruction. Recent work has further linked hidden states to competing interlayer stacking configurations \cite{liu2025nonvolatile}. Although the present measurements do not directly identify the structural origin of the HPLMS, exfoliation likely modify the energetic balance between competing insulating and metallic states through changes in stacking, interlayer coupling, or cooling dynamics, thereby enabling spontaneous stabilization of a hidden-phase-like electronic structure under equilibrium conditions. Although metallic surface electronic structures have been reported in related TaS$_2$ systems \cite{gofman2025chiral,date2026charge}, the spectroscopic signatures observed here are more naturally interpreted within the framework of hidden-state physics than as a simple surface-derived metallic state or rigid chemical-potential shift.

The spontaneous appearance of the HPLMS demonstrates that this hidden-phase-like electronic configuration represents a viable equilibrium state rather than a purely transient one. More broadly, these results establish a spectroscopic benchmark for understanding metastable phases in correlated materials and suggest that dimensional engineering and controlled sample preparation provide an alternative pathway for accessing hidden-state physics without optical or electrical switching. Such an approach offers new opportunities for tailoring the electronic structure and transport properties of 1T-TaS$_2$ and related layered quantum materials.

\section*{MATERIALS AND METHODS}

\subsection*{Samples} Single crystals of 1T-TaS\textsubscript{2} were obtained from 2Dsemiconductors. High-quality 1T-TaS$_2$ single crystals were exfoliated onto conducting substrates in an inert environment. Flakes with a range of thicknesses were selected for ARPES measurements. Approximate thicknesses were inferred from exfoliation conditions, optical contrast, and relative yield statistics. More details of the flake preparation is given in the Supplementary Information.

\subsection*{ARPES experiments}

$\mu$-ARPES measurements were carried out at the ESM beamline (21-ID-1) of NSLS-II~\cite{rajapitamahuni2024electron}, employing a DA30 Scienta electron spectrometer. The samples were cleaved \textit{in situ}, and the base pressure in the photoemission chambers was maintained below $3 \times 10^{-11}$~Torr. At 21-ID-1, the incident photon beam was focused to a spot size of approximately $5~\mu\mathrm{m}^2$, with synchrotron radiation impinging on the sample surface at an angle of $55^\circ$. The out-of-plane momentum component $k_z$ was determined using the free-electron final-state approximation~\cite{damascelli2003angle}: $k_z = \sqrt{\frac{2m}{\hbar^2}\left(E_{\mathrm{kin}}\cos^2\theta + V_0\right)}$, where $E_{\mathrm{kin}}$ denotes the kinetic energy of the photoelectrons, $\theta$ is the emission angle relative to the surface normal, $V_0$ is the inner potential (to be 17.5~eV for 1T-TaS$_2$~\cite{wang2020band}), $m$ is the free-electron mass, and $\hbar$ is the reduced Planck constant.

\subsection*{Acknowledgments}

This research used resources at the ESM (21-ID-1) beamline of the National Synchrotron Light Source II, a U.S. Department of Energy (DOE) Office of Science User Facility operated by Brookhaven National Laboratory under Contract No.~DE-SC0012704. This work also used the theory and computation resources and the Quantum Material Press (QPress) of the Center for Functional Nanomaterials, which is a U.S. DOE Office of Science User Facility at the Brookhaven National Laboratory under Contract number DE-SC0012704.). Author T.Y. notes that this research benefited from support provided by the Xiamen University Malaysia research grant (Grant No.~IPHY/0008).

\subsection*{Data availability} The data that support the findings of this study are available from the corresponding author upon request.

\subsection*{Competing Interests} The authors declare no competing interests.

\bibliography{references}

\end{document}